\documentclass[twocolumn,english,prd]{revtex4-2}
\usepackage[T1]{fontenc}
\usepackage[latin9]{inputenc}
\usepackage{amsmath}
\usepackage{amssymb}

\makeatletter
\usepackage{bm}
\usepackage{amsthm}
\usepackage[]{sidecap}
\sidecaptionvpos{figure}{c}
\usepackage{amssymb}
\usepackage{lipsum}

\makeatother

\usepackage{babel}
\begin{document}
\title{Causal Finite-Tick Dynamics as a Modern Resolution of the Classical
Radiation Reaction Problem}
\author{Hadi Z. Olyaei}
\address{Robert Bosch GmbH, Abstatt, Germany}
\begin{abstract}
The radiation--reaction problem in classical electrodynamics, embodied
in the Abraham--Lorentz--Dirac (ALD) equation, has long resisted
a causal and stable formulation. The ALD dynamics admits runaway and
pre-accelerating solutions, while the Landau--Lifshitz (LL) equation,
though empirically successful, relies on an order-reduction procedure
whose physical mechanism remains opaque. In this work we show that
a strictly causal, finite-difference formulation of the particle equations
of motion---termed \emph{Tick--Tock (TT) dynamics}---naturally
regularizes the self-force and resolves the classical pathologies
of radiation reaction. The TT update law evolves the Lorentz acceleration
through discrete proper-time intervals of duration $\Delta t$, ensuring
that radiation recoil depends only on past motion. Energy--momentum
conservation is exact at each step: the radiated power appears explicitly
as a Larmor term, while the Schott-like contribution telescopes to
boundary terms, removing the need for hidden near-field energy reservoirs.
The discrete evolution operator is manifestly Lorentz covariant, bounded
at all frequencies, and reduces continuously to the LL equation in
the limit $\Delta t\to0$. Identifying $\Delta t$ with the classical
radiation--reaction timescale $\tau_{0}$ introduces a natural causal
cutoff near the onset of quantum electrodynamical corrections. The
resulting framework provides a mathematically consistent and physically
transparent resolution of the radiation--reaction problem, reconciling
causality, stability, and energy balance within a single discrete-time
formalism.
\end{abstract}
\maketitle

\section{Introduction}

The radiation--reaction problem has remained one of the most persistent
open questions in classical electrodynamics. Whenever a charged particle
is accelerated, it emits electromagnetic radiation, and conservation
of energy--momentum requires that a recoil force act back on the
particle. A consistent equation of motion that captures this recoil
has, however, proven elusive for more than a century. The earliest
formulations, due to Abraham and Lorentz, led to the well-known Abraham--Lorentz
(AL) equation. Dirac subsequently extended this to a fully covariant
form, now known as the Abraham--Lorentz--Dirac (ALD) equation \citep{PAMDirac1938}.
While conceptually appealing, the ALD dynamics exhibits two notorious
pathologies: \emph{runaway solutions}, in which a particle accelerates
indefinitely without external force, and \emph{pre-acceleration},
in which the charge responds before the applied force is present.
These features are widely regarded as unphysical, and they cast doubt
on the adequacy of ALD as a fundamental description \citep{Rohrlich2007,Griffiths2012,Klepikov1985,Spohn2004}. 

A pragmatic alternative was developed by Landau and Lifshitz (LL)
\citep{Landau1989}, who eliminated the problematic higher-derivative
term by reducing the order of the ALD equation. Their prescription
replaces the derivative of the acceleration with the derivative of
the Lorentz force, resulting in a second-order law that avoids both
runaways and pre-acceleration. The LL equation has since become the
standard tool in plasma and laser physics \citep{DiPiazza2012,Gonoskov2022,Kravets2013},
and is consistent with all experimental tests so far, provided that
radiation reaction is a small perturbation to the external force.
Nevertheless, the LL equation is not derived from first principles,
but rather as an approximation, and its physical interpretation relies
on the introduction of the so-called Schott energy. This near-field
energy acts as a hidden reservoir that ensures overall conservation,
but it lacks direct observability and leaves the physical mechanism
obscure \citep{Spohn2000,Ford1991}. 

The limitations of ALD and LL have motivated a wide variety of alternative
approaches. Extended-charge models attempt to regularize the self-field
by smearing out the point charge, but at the cost of introducing ad
hoc parameters \citep{Ford1993,Medina2006,Bauer2013}. Caldirola's
``chronon'' theory \citep{Caldirola1956} introduced a fundamental
discrete time step in the electron's motion, but without a transparent
accounting of energy--momentum balance. Stochastic electrodynamics
\citep{Krivitskii1991}, renormalization techniques \citep{Sokolov2009_strongfield},
and effective field-theory treatments \citep{DiPiazza2008,Ilderton2013,Torgrimsson2020,Spohn2004}
have each addressed aspects of the problem, yet none has provided
a simple, deterministic, and causal update law that avoids pathologies
while maintaining exact energy conservation at each step. 

In this work we demonstrate such a formulation. We introduce \emph{Tick--Tock (TT) dynamics},
a finite-tick description in which the motion of a charged particle
proceeds in discrete updates of duration $\Delta t$. In this framework,
radiation recoil arises from finite-difference changes in the Lorentz
acceleration, rather than from evaluating the singular self-field
of the particle. The TT scheme ensures strict causality (depending
only on present and past states), eliminates runaway solutions, and
makes the energy balance transparent: the Larmor radiation loss appears
explicitly, while the Schott-like contribution reduces to a telescoping
boundary term. In the continuum limit $\Delta t\to0$, TT reproduces
the LL equation, thereby guaranteeing agreement with all present experiments,
while at the same time extending its applicability to regimes where
external fields vary rapidly. 

Beyond resolving the pathologies of ALD and clarifying the foundations
of LL, TT dynamics introduces a natural high-frequency cutoff, suppressing
unphysical growth in discontinuous fields. This suggests that the
characteristic radiation--reaction timescale $\tau_{0}$ may serve
as a fundamental unit of dynamical updates, potentially marking the
boundary between classical and quantum domains \citep{Higuchi2002,Niel2018,Dunne2009}.

The remainder of this article develops the TT formalism in detail.
Section (\ref{sec:background}) reviews the classical background of
radiation reaction and the limitations of existing formulations. Section
(\ref{sec:TT}) introduces the non-relativistic TT update law, derives
its energy balance, and examines its behavior under constant and discontinuous
forces. Section (\ref{sec:covariant}) presents the covariant extension
and proves strict causality, orthogonality, and energy--momentum
conservation. Section (\ref{sec:discussion}) discusses the physical
implications, including connections to effective field theory and
quantum electrodynamics, and Section (\ref{sec:Outlook}) offers an
outlook on future applications. 

\paragraph{Conventions.---}

We work in SI units in Secs. \ref{sec:background}--\ref{sec:TT}
(with $c$ explicit) and adopt natural units $c=1$ in Sec. \ref{sec:covariant}
onward. An overdot denotes $d/dt$, while a dot over a four-vector
denotes $d/d\tau$. Bold symbols are $3$-vectors; Greek indices are
spacetime components with metric $g=\mathrm{diag}(1,-1,-1,-1)$.

\section{Theoretical Background\label{sec:background}}

\subsection{Abraham--Lorentz and Abraham--Lorentz--Dirac equations}

The first systematic attempts to describe radiation reaction date
back to Abraham and Lorentz, who proposed that the self-force acting
on an accelerated point charge can be written as

\begin{equation}
m\mathbf{a}=\mathbf{F}_{\mathrm{ext}}+m\tau_{0}\dot{\mathbf{a}},\label{eq:AL}
\end{equation}
where $\mathbf{a}$ is the particle acceleration, $\mathbf{F}_{\mathrm{ext}}$
the external force, and 

\begin{equation}
\tau_{0}=\frac{2}{3}\frac{q^{2}}{4\pi\epsilon_{0}mc^{3}}\label{eq:tau0}
\end{equation}
is the characteristic radiation--reaction timescale \citep{Rohrlich2007,Jackson2021}.
Equation (\ref{eq:AL}) immediately predicts unphysical phenomena:
in the absence of any external force, solutions may grow exponentially,
leading to \emph{runaway acceleration}. Moreover, because the equation
involves the derivative of acceleration, it admits solutions in which
the particle accelerates \emph{before} the force is applied, violating
causality. Dirac reformulated the problem in a manifestly covariant
way, obtaining what is now called the Abraham--Lorentz--Dirac (ALD)
equation \citep{PAMDirac1938}:

\begin{equation}
m\frac{du^{\mu}}{d\tau}=qF^{\mu\nu}u_{\nu}+m\tau_{0}\left(\frac{d^{2}u^{\mu}}{d\tau^{2}}+\frac{du_{\nu}}{d\tau}\frac{du^{\nu}}{d\tau}u^{\mu}\right),\label{eq:ALD}
\end{equation}
where $u^{\mu}$ is the four-velocity, $\tau$ the proper time, and
$F^{\mu\nu}$ the external electromagnetic field tensor. Equation
(\ref{eq:ALD}) reproduces the same unphysical behaviors as Eq. (\ref{eq:AL}),
and has been the subject of extensive debate and reinterpretation
\citep{Spohn2004,Gralla2009,Bauer2013,Bild2019}. 

\subsection{Landau--Lifshitz reduction of order}

A pragmatic resolution was introduced by Landau and Lifshitz (LL)
\citep{Landau1989}, who noted that radiation reaction typically constitutes
only a small perturbation to the motion induced by the external force.
They therefore proposed to \emph{reduce the order} of Eq. (\ref{eq:ALD})
by replacing the derivative of the acceleration with the derivative
of the Lorentz force,

\begin{equation}
\dot{\mathbf{a}}\;\rightarrow\;\frac{1}{m}\frac{d\mathbf{F}_{\mathrm{ext}}}{dt}.
\end{equation}
The resulting second-order equation avoids both runaway and pre-accelerating
solutions, and has become the standard equation of motion in plasma
and laser physics \citep{DiPiazza2012,Kravets2013}, see also recent
comprehensive reviews \citep{Gonoskov2022}. In covariant form, the
LL update reads

\begin{multline}
m\frac{du^{\mu}}{d\tau}=qF^{\mu\nu}u_{\nu}+q\tau_{0}\Big(\partial_{\alpha}F^{\mu\nu}u^{\alpha}u_{\nu}\\
+\frac{q}{m}F^{\mu\nu}F_{\nu\rho}u^{\rho}-\frac{q}{m}(F_{\alpha\nu}u^{\nu})(F^{\alpha\rho}u_{\rho})u^{\mu}\Big).
\end{multline}
 This formulation is well-behaved in all experimentally accessible
regimes, but it comes at the cost of introducing a hidden bookkeeping
device: the \emph{Schott energy}. 

\paragraph{The role of Schott energy.---}

In both the ALD and LL descriptions, energy conservation is enforced
through a subtle interplay between radiated power and a near-field
term, often called the Schott energy \citep{Spohn2000,Ford1991}.
For a non-relativistic particle, the Larmor formula for the radiated
power is

\begin{equation}
P_{\mathrm{rad}}=\frac{q^{2}a^{2}}{6\pi\epsilon_{0}c^{3}},
\end{equation}
where $a^{2}=\mathbf{a}\cdot\mathbf{a}$. In the AL formulation, the
work done by the self-force can be decomposed into an instantaneous
radiation term $-m\tau_{0}a^{2}$ and a total derivative of the Schott
energy. In LL theory, the Schott term persists as a ``hidden reservoir''
that ensures global energy--momentum balance, but it is not directly
observable. This reliance on a hidden energy term is often regarded
as unsatisfactory: it obscures the local mechanism by which radiation
recoil operates, and complicates numerical implementations in discontinuous
or rapidly varying fields \citep{Kirk2009,Yaghjian1992,Liseykina2016}. 

\subsection{Open issues}

Despite the wide use of LL, the absence of a strictly causal, deterministic,
and transparent update law for radiation reaction continues to motivate
theoretical investigation. Extended-charge models \citep{Ford1993},
stochastic electrodynamics \citep{Krivitskii1991}, and renormalization-based
approaches \citep{Spohn2004} have each offered valuable insight,
but none has provided a fully satisfactory resolution. In this context,
discrete-time formulations are especially intriguing. Caldirola's
``chronon'' model \citep{Caldirola1956} suggested introducing a
fixed time step into the dynamics, but subsequent analysis found the
approach essentially ad hoc, with energy--momentum balance left unclear
\citep{DiPiazza2012}, see also detailed critique in \citep{Burton2014}.
These challenges motivate the development of an alternative formulation
in which causality, stability, and energy conservation are built in
at the fundamental level. In the following sections we show such a
framework: \emph{Tick--Tock dynamics}, a finite-tick approach that
recovers LL in the continuum limit while eliminating the pathologies
of ALD and clarifying the role of Schott energy.

\section{Tick--Tock Dynamics Framework\label{sec:TT}}

\subsection{Motivation and discrete-time formulation}

We show that the motion of a charged particle advances in discrete
steps of duration $\Delta t$, which we term \emph{ticks}. The particle
state at tick $n\in\mathbb{N}_{\geq0}$ is specified by $(\mathbf{x}_{n},\mathbf{v}_{n})$
at time $t_{n}=n\Delta t$. At each tick, the external Lorentz acceleration
is evaluated as

\begin{equation}
\mathbf{a}_{n}^{L}=\frac{1}{m}\,\mathbf{F}_{\mathrm{ext}}(\mathbf{x}_{n},\mathbf{v}_{n},t_{n}).\label{eq:LorentzAcc}
\end{equation}
From this we define the \emph{discrete jerk} as the finite difference

\begin{equation}
\mathbf{j}_{n}=\frac{\mathbf{a}_{n}^{L}-\mathbf{a}_{n-1}^{L}}{\Delta t}.\label{eq:jerk}
\end{equation}
The TT update law for the actual acceleration is then postulated to
be

\begin{equation}
\mathbf{a}_{n}=\mathbf{a}_{n}^{L}+\tau_{0}\,\mathbf{j}_{n},\label{eq:TTupdate}
\end{equation}
where $\tau_{0}$ is the classical radiation--reaction timescale
defined in Eq. (\ref{eq:tau0}). Velocities and positions are subsequently
updated by finite-difference integration. Physically, Eq. (\ref{eq:TTupdate})
states that radiation reaction arises not from evaluating the singular
self-field of the particle, but from causal \emph{tick-to-tick changes}
in the emission pattern. Each update involves only present and past
ticks, ensuring strict causality, while the finite-difference form
eliminates the runaway instabilities associated with the ALD equation.

\subsection{Derivation from an extended-charge model}

The TT law can be derived heuristically by considering a rigid, uniformly
charged sphere of radius $r_{0}$ and total charge $q$ centered at
$\mathbf{x}(t)$. Expanding the retarded Lorentz acceleration at time
$t_{n}-R/c$ to first order in $R/c$, with $0\le R\le r_{0}$, and
replacing the derivative by a finite difference yields

\begin{equation}
\mathbf{a}^{L}(t_{n}-R/c)\approx\mathbf{a}_{n}^{L}-\frac{R}{c}\,\frac{\mathbf{a}_{n}^{L}-\mathbf{a}_{n-1}^{L}}{\Delta t}.
\end{equation}
Integrating the retarded field of the charge distribution to order
$1/c^{3}$ gives the self-force

\begin{equation}
\mathbf{F}_{n}^{\mathrm{self}}=-\frac{2}{3}\frac{q^{2}}{4\pi\epsilon_{0}c^{2}r_{0}}\,\mathbf{a}_{n}+\frac{2}{3}\frac{q^{2}}{4\pi\epsilon_{0}c^{3}}\frac{\mathbf{a}_{n}^{L}-\mathbf{a}_{n-1}^{L}}{\Delta t}.
\end{equation}
Here $m_{\mathrm{em}}=\frac{2}{3}\,\frac{q^{2}}{4\pi\epsilon_{0}c^{2}r_{0}}$
is the electromagnetic contribution to the mass, and $m\tau_{0}=\frac{2}{3}\,\frac{q^{2}}{4\pi\epsilon_{0}c^{3}},$
so the finite term equals $m\tau_{0}\,\frac{\mathbf{a}_{n}^{L}-\mathbf{a}_{n-1}^{L}}{\Delta t},$
consistent with the discrete analogue of the Abraham--Lorentz self-force.
The first, divergent term can be absorbed into a renormalized mass,
while the second yields the finite radiation--reaction contribution.
This leads directly to Eq. (\ref{eq:TTupdate}). 

\subsection{Energy balance per tick \label{subsec:TTenergy}}

The TT update law admits a transparent and fully causal decomposition
of the energy balance. Over one tick, the change in kinetic energy
is

\begin{equation}
\Delta K_{n}\simeq m\,\mathbf{v}_{n-\tfrac{1}{2}}\!\cdot\!\mathbf{a}_{n}\,\Delta t,
\end{equation}
where $\mathbf{v}_{n-\tfrac{1}{2}}\equiv(\mathbf{v}_{n}+\mathbf{v}_{n-1})/2$
is the midpoint velocity. Substituting the TT update

\begin{equation}
\mathbf{a}_{n}=\mathbf{a}_{n}^{L}+\tau_{0}\,\frac{\mathbf{a}_{n}^{L}-\mathbf{a}_{n-1}^{L}}{\Delta t}\label{eq:TTupdate_reuse}
\end{equation}
yields

\begin{equation}
\Delta K_{n}\approx\mathbf{F}_{\mathrm{ext},n}\!\cdot\!\mathbf{v}_{n-\tfrac{1}{2}}\,\Delta t+m\tau_{0}\,\mathbf{v}_{n-\tfrac{1}{2}}\!\cdot\!\frac{\mathbf{a}_{n}^{L}-\mathbf{a}_{n-1}^{L}}{\Delta t}\,\Delta t.\label{eq:Kchange_raw}
\end{equation}
To evaluate the second term, we employ a discrete integration-by-parts
identity (valid up to $\mathcal{O}(\Delta t)$) \citep{Baylis2002},
which is the finite-difference analogue of the continuum relation
$\mathbf{v}\!\cdot\!\dot{\mathbf{a}}=\tfrac{d}{dt}(\mathbf{v}\!\cdot\!\mathbf{a})-a^{2}$:

\begin{equation}
\mathbf{v}_{n-\tfrac{1}{2}}\!\cdot\!\big(\mathbf{a}_{n}^{L}-\mathbf{a}_{n-1}^{L}\big)=\big[(\mathbf{v}\!\cdot\!\mathbf{a}^{L})_{n}-(\mathbf{v}\!\cdot\!\mathbf{a}^{L})_{n-1}\big]-\big(\mathbf{a}_{n-\tfrac{1}{2}}^{L}\big)^{2}\,\Delta t+\mathcal{O}(\Delta t^{2}).\label{eq:discreteIBP}
\end{equation}

Here $\mathbf{a}_{n-\tfrac{1}{2}}^{L}\!\equiv(\mathbf{a}_{n}^{L}+\mathbf{a}_{n-1}^{L})/2$.
Substituting Eq. (\ref{eq:discreteIBP}) into Eq. (\ref{eq:Kchange_raw})
gives, to leading order in $\tau_{0}$,

\begin{align}
\Delta K_{n} & \approx\mathbf{F}_{\mathrm{ext},n}\!\cdot\!\mathbf{v}_{n-\tfrac{1}{2}}\,\Delta t-m\tau_{0}\,\big(\mathbf{a}_{n-\tfrac{1}{2}}^{L}\big)^{2}\,\Delta t+\Delta S_{n}\label{eq:TTenergy}\\[4pt]
 & +\mathcal{O}(\tau_{0}\Delta t^{2}),\nonumber \\
S_{n} & \equiv m\tau_{0}\,\mathbf{v}_{n}\!\cdot\!\mathbf{a}_{n}^{L},\qquad\Delta S_{n}=S_{n}-S_{n-1}.
\end{align}
Equation (\ref{eq:TTenergy}) separates the tick-wise energy balance
into three transparent contributions: The first term represents the
work performed by the external field, the second corresponds to the
instantaneous Larmor radiation loss $-m\tau_{0}(\mathbf{a}_{n-\tfrac{1}{2}}^{L})^{2}\Delta t$,
and the third, $\Delta S_{n}$, is a discrete Schott-like boundary
term that telescopes across ticks. Because the Schott contribution
collapses to endpoints when summed, no hidden energy reservoir is
required. Over many ticks, the average loss reproduces the classical
Larmor formula, while the energy exchange is accounted for explicitly
at each update. For a constant external force ($\mathbf{a}_{n}^{L}=\mathbf{a}_{n-1}^{L}=\mathbf{a}^{L}$),
the increment $\Delta S_{n}=m\tau_{0}(\mathbf{a}^{L})^{2}\Delta t$
exactly cancels the instantaneous Larmor loss, so that $\Delta K_{n}=\mathbf{F}_{\mathrm{ext}}\!\cdot\!\mathbf{v}_{n-\tfrac{1}{2}}\Delta t$
and $\mathbf{a}_{n}=\mathbf{a}^{L}$. Thus, under steady forcing,
the TT dynamics are strictly causal, energy-conserving, and lossless.
This special case is examined again in the next sections, where its
dynamical and spectral implications are discussed.

\subsection{Constant-force case}

For a constant external (Lorentz) force, the acceleration $a_{n}^{L}$
is time independent. In this case, the finite-difference term in the
TT update,

\[
a_{n}=a_{n}^{L}+\tau_{0}\frac{a_{n}^{L}-a_{n-1}^{L}}{\Delta t},
\]
vanishes identically, since $a_{n}^{L}=a_{n-1}^{L}=a_{L}$. Thus the
TT and LL formulations coincide,

\[
a_{n}=a_{L},
\]
and the particle experiences a uniform acceleration equal to the Lorentz
value. The radiation-reaction correction is therefore zero in steady
conditions. Although this result appears trivial, it has clear physical
significance. Because the forcing is constant, the finite-difference
operator contributes no residual phase shift or damping, confirming
that the finite-tick update preserves equilibrium in static fields.
The particle\textquoteright s self-force does not accumulate any artificial
memory from past ticks, demonstrating that the TT formulation is strictly
causal yet lossless for steady motion. In spectral terms, the constant-force
limit corresponds to the zero-frequency point, where $K_{\mathrm{TT}}(0)=K_{\mathrm{LL}}(0)=1$.
This anchors the entire TT response at its correct zero-frequency
value, ensuring that only genuinely time-varying components of the
motion are modified by the finite-tick kernel. Physically, this means
that the TT model leaves uniform acceleration and the corresponding
constant radiation power unchanged, while selectively regularizing
the self-interaction at higher frequencies. Hence, the constant-force
case verifies that the TT dynamics reproduce the correct static limit
and introduce no spurious effects in slowly varying fields.

\subsection{Spectral properties and sharp-edge response}

Having established that the TT formulation reproduces the correct
static limit under constant forcing, we now examine its behaviour
for time-dependent or discontinuous external fields. The frequency-domain
perspective reveals how the discrete dynamics regularize the high-frequency
pathologies of the LL model. Applying the discrete-time Fourier transform
to the update

\begin{equation}
a_{n}=a_{n}^{L}+\tau_{0}\frac{a_{n}^{L}-a_{n-1}^{L}}{\Delta t},
\end{equation}
one obtains the kernels

\begin{equation}
K_{\mathrm{TT}}(\omega)=1+i\omega\tau_{0}\,e^{-i\omega\Delta t/2}\operatorname{sinc}\!\Big(\frac{\omega\Delta t}{2}\Big),\quad K_{\mathrm{LL}}(\omega)=1+i\omega\tau_{0},
\end{equation}
where $\operatorname{sinc}x=\sin x/x$ and$\Omega=\omega\Delta t$.
Only the radiation-reaction term $i\omega\tau_{0}$ is modified by
the finite difference; the Lorentz response (unity term) remains unchanged.
The factor $e^{-i\omega\Delta t/2}\operatorname{sinc}(\omega\Delta t/2)$
introduces both a phase delay and a smooth attenuation that limit
the growth of the self-reaction term at high frequency. Consequently,
the TT kernel remains bounded for all $|\omega\Delta t|\le\pi$, whereas
the LL kernel grows linearly with frequency, $|K_{\mathrm{LL}}|\propto\omega\tau_{0}$.
For $\omega\Delta t\ll1$, the two kernels coincide to leading order,

\begin{equation}
K_{\mathrm{TT}}=K_{\mathrm{LL}}+O(\omega^{2}\tau_{0}\Delta t),
\end{equation}
so the finite-tick dynamics reduce smoothly to the LL limit for slowly
varying fields.

\subsubsection{Significance of the Finite-Tick Dynamics for Sharp-Edge Responses}

Discontinuous or rapidly varying external forces pose the most severe
challenge to any radiation-reaction formulation. In the classical
ALD and LL descriptions, the radiation-reaction term involves the
time derivative of the acceleration, which couples the self-force
instantaneously to the highest temporal frequencies of the motion.
This derivative form makes the LL operator unbounded in frequency
space: high-frequency components of the Lorentz acceleration $a_{L}(t)$
are amplified without limit as $|K_{\mathrm{LL}}|^{2}=1+(\omega\tau_{0})^{2}\rightarrow\infty$.
Such behaviour underlies both the numerical stiffness of the LL equation
and its physical pathologies in the presence of sharp edges or impulsive
driving. The TT formulation replaces this instantaneous derivative
by a \emph{finite causal difference}. Each tick of duration $\Delta t$
connects the current acceleration to its immediate past, $a_{n-1}$,
rather than to an infinitesimal neighbour. In the frequency domain
this operation introduces the factor

\begin{equation}
e^{-i\omega\Delta t/2}\operatorname{sinc}\!\Big(\frac{\omega\Delta t}{2}\Big),
\end{equation}
so that the radiation-reaction component becomes

\begin{equation}
\delta a_{\mathrm{TT}}(\omega)=i\omega\tau_{0}\,e^{-i\omega\Delta t/2}\operatorname{sinc}\!\Big(\frac{\omega\Delta t}{2}\Big)\,a_{L}(\omega).
\end{equation}
Because $|\operatorname{sinc}(x)|\le1$ for all $x$, the TT operator
is \emph{bounded}: for every frequency within the discrete spectrum
$|\omega\Delta t|\le\pi$,

\begin{equation}
|\delta a_{\mathrm{TT}}(\omega)|\le\frac{2\tau_{0}}{\Delta t}\,|a_{L}(\omega)|.
\end{equation}
This follows from $|\Omega\,\operatorname{sinc}(\Omega/2)|=2|\sin(\Omega/2)|\le2$.
Hence the finite-tick formulation can at most amplify the Lorentz
acceleration by a factor of $2\tau_{0}/\Delta t$, regardless of frequency.
In contrast, the LL derivative operator has no such upper bound---its
gain increases indefinitely with $\omega$. This demonstrates in explicit
mathematical form that the TT dynamics \emph{damp} the radiation-reaction
response at every frequency. As a result, the TT scheme remains stable
and well behaved even when the driving field changes abruptly. Physically,
this boundedness has a transparent meaning. The finite-tick self-interaction
implies that the particle cannot react to its own radiation field
faster than one tick. Sharp edges in the external force therefore
generate smoothed self-forces rather than instantaneous, infinite
jerks. The phase factor $e^{-i\omega\Delta t/2}$ introduces a half-tick
retardation, ensuring strict causality: the radiation reaction always
lags the external field by a finite interval. The practical consequence
is that the TT power spectrum

\begin{equation}
P_{\mathrm{TT}}(\omega)=|K_{\mathrm{TT}}(\omega)|^{2}=\big|1+i\omega\tau_{0}\,e^{-i\omega\Delta t/2}\operatorname{sinc}(\omega\Delta t/2)\big|^{2},
\end{equation}
remains bounded as $\omega\!\to\!\pi/\Delta t$, whereas

\begin{equation}
P_{\mathrm{LL}}(\omega)=|K_{\mathrm{LL}}(\omega)|^{2}=1+(\omega\tau_{0})^{2}
\end{equation}
which diverges quadratically. The contrast between the LL and TT formulations
becomes clear in the frequency domain. While the LL spectrum grows
without limit at high frequencies, the TT spectrum remains bounded
and \emph{saturates} near the discrete Nyquist frequency. Using the
discrete-time relation

\[
K_{\mathrm{TT}}(\Omega)=1+\frac{\tau_{0}}{\Delta t}\big(1-e^{-i\Omega}\big),\qquad\Omega\equiv\omega\Delta t,
\]
the total gain becomes

\begin{align*}
|K_{\mathrm{TT}}(\Omega)|^{2} & =\big[1+\tfrac{\tau_{0}}{\Delta t}(1-\cos\Omega)\big]^{2}+\big[\tfrac{\tau_{0}}{\Delta t}\sin\Omega\big]^{2}\\
 & =1+4\frac{\tau_{0}}{\Delta t}\!\left(1+\frac{\tau_{0}}{\Delta t}\right)\sin^{2}\!\frac{\Omega}{2}.
\end{align*}
This expression remains finite for all $\Omega$; its maximum value
is

\[
|K_{\mathrm{TT}}(\pi)|^{2}=(1+2\tfrac{\tau_{0}}{\Delta t})^{2},
\]
so the magnitude of the TT kernel satisfies the bound

\[
|K_{\mathrm{TT}}(\Omega)|\le1+2\,\tfrac{\tau_{0}}{\Delta t}.
\]
This boundedness guarantees that the TT formulation does not grow
without bound with frequency, in contrast to the LL kernel whose magnitude
increases linearly as $|K_{\mathrm{LL}}|\sim\omega\tau_{0}$.

At the Nyquist limit $\Omega=\pi$, the magnitude saturates to $|K_{\mathrm{TT}}(\pi)|^{2}=(1+2\tau_{0}/\Delta t)^{2}$,
in contrast to the unbounded growth $|K_{\mathrm{LL}}|^{2}=1+(\omega\tau_{0})^{2}$
of the LL kernel. Thus, the TT formulation acts as a \emph{causal high-frequency regularization}:
it suppresses unphysical amplification at large frequencies without
altering the low-frequency limit. For $\omega\Delta t\ll1$, the expansion

\[
K_{\mathrm{TT}}=1+i\omega\tau_{0}+\mathcal{O}(\omega^{2}\tau_{0}\Delta t)
\]
shows that TT reduces smoothly to the LL form, preserving the correct
continuum behaviour while ensuring bounded, stable response across
the entire discrete spectrum. Finally, if one identifies the tick
interval with the characteristic radiation-reaction time, $\Delta t\simeq\tau_{0}$,
the TT cutoff frequency $\omega_{\mathrm{Ny}}=\pi/\Delta t$ lies
near a few hundred\textasciitilde MeV for the electron, close to
the energy scale where quantum-electrodynamic processes begin to dominate.
This numerical coincidence suggests a natural physical interpretation:
the TT scheme regularizes the classical theory precisely at the boundary
where a quantum description becomes essential. While this correspondence
is heuristic rather than derived, it highlights the broader conceptual
significance of the finite-tick approach---a \emph{causal, bounded, and physically motivated regularization}
of classical radiation reaction.

\paragraph{Summary.---}

The finite-tick formulation introduces a discrete, causal delay that
ensures the self-reaction depends on past motion by a finite interval.
Its response remains bounded at all frequencies, in contrast to the
unbounded growth of the LL operator, and therefore smooths any sharp-edge
forcing into finite, continuous self-forces. In the limit of smooth
fields ($\omega\Delta t\ll1$), the TT dynamics reduce to the LL form,
recovering the correct continuum behaviour. Setting $\Delta t\simeq\tau_{0}$
places the natural cutoff frequency near the QED transition scale,
suggesting that the TT scheme provides a physically motivated and
causal regularization of classical radiation reaction.

\subsection{Continuum limit and recovery of Landau--Lifshitz}

As $\Delta t\to0$, the discrete jerk approaches the time derivative
of the Lorentz acceleration,

\begin{equation}
\lim_{\Delta t\to0}\frac{\mathbf{a}_{n}^{L}-\mathbf{a}_{n-1}^{L}}{\Delta t}=\dot{\mathbf{a}}^{L}(t).
\end{equation}
Equation (\ref{eq:TTupdate}) then reduces to

\begin{equation}
\mathbf{a}(t)=\mathbf{a}^{L}(t)+\tau_{0}\,\dot{\mathbf{a}}^{L}(t),
\end{equation}
which is precisely the Landau--Lifshitz form to leading order in
$\tau_{0}$. Thus TT is consistent with all known experimental tests
of radiation reaction, while extending LL into regimes where fields
are discontinuous or vary on timescales comparable to $\tau_{0}$.

\section{Covariant Formulation\label{sec:covariant}}

\subsection{Proper-time discretization \label{subsec:4.1 Proper-time-discretization}}

To formulate TT dynamics covariantly, we parametrize the particle
worldline by proper time $\tau$ with metric signature $g=\mathrm{diag}(1,-1,-1,-1)$.
Note that $\tau_{0}$ retains its meaning as the classical radiation--reaction
timescale defined in (\ref{eq:tau0}); the symbol $\tau_{\mathrm{i}}$
is used below for the initial proper time. Throughout, the Minkowski
inner product is

\[
A\!\cdot B\equiv g_{\mu\nu}A^{\mu}B^{\nu},
\]
so that timelike four-velocities satisfy $u^{\mu}u_{\mu}=1$ and the
Lorentz four-acceleration obeys $A^{L}\!\cdot A^{L}<0$. The four-velocity
is defined as

\begin{equation}
u^{\mu}=\frac{dx^{\mu}}{d\tau},\qquad u^{\mu}u_{\mu}=1,
\end{equation}
and the four-acceleration is

\begin{equation}
A^{\mu}=\frac{du^{\mu}}{d\tau}.
\end{equation}
Discrete ticks are separated by a proper-time interval $\Delta\tau$,
so that

\[
\tau_{n}=\tau_{\mathrm{i}}+n\,\Delta\tau.
\]
At each tick, the external Lorentz four-acceleration is

\begin{equation}
A_{L,n}^{\mu}=\frac{q}{m}\,F^{\mu\nu}(x_{n})u_{\nu,n},\label{eq:Lorentz4acc}
\end{equation}
which satisfies the orthogonality condition $u_{n,\mu}A_{L,n}^{\mu}=0$. 

\subsection{Projector and TT update law}

To ensure that the full four-acceleration remains orthogonal to the
four-velocity, we introduce the projection operator

\begin{equation}
\Pi^{\mu\nu}(u_{n})=g^{\mu\nu}-u_{n}^{\mu}u_{n}^{\nu}.
\end{equation}
The TT update rule in covariant form is then

\begin{equation}
A_{n}^{\mu}=\Pi^{\mu\nu}(u_{n})\left[A_{\nu,n}^{L}+\tau_{0}\frac{A_{\nu,n}^{L}-A_{\nu,n-1}^{L}}{\Delta\tau}\right].\label{eq:TTcovariant}
\end{equation}
The velocity and position are updated by

\begin{align}
u_{n+1}^{\mu} & =u_{n}^{\mu}+A_{n}^{\mu}\Delta\tau,\\
x_{n+1}^{\mu} & =x_{n}^{\mu}+u_{n+1}^{\mu}\Delta\tau.
\end{align}
Finally, a renormalization step ensures norm preservation,

\begin{equation}
u_{n+1}^{\mu}\;\mapsto\;\frac{u_{n+1}^{\mu}}{\sqrt{u_{n+1}^{\alpha}u_{n+1,\alpha}}}.
\end{equation}

\subsection{Properties of the covariant TT dynamics}

We now summarize the key structural properties of Eq. (\ref{eq:TTcovariant}). 

\paragraph{Lorentz covariance.---}

Every ingredient in Eq. (\ref{eq:TTcovariant}) is a tensor constructed
from covariant quantities ($F^{\mu\nu}(x_{n})$, $u_{n}^{\mu}$, $g^{\mu\nu}$,
and $\Delta\tau$), ensuring that the update law is manifestly Lorentz
covariant.

\paragraph{Orthogonality.---}

By construction, $\Pi^{\mu\nu}(u_{n})u_{n,\mu}=0$, so that $u_{n,\mu}A_{n}^{\mu}=0$.
Thus the acceleration is always orthogonal to the velocity, preserving
the unit-norm condition for $u^{\mu}$.

\paragraph{Norm preservation.---}

From the update $u_{n+1}^{\mu}=u_{n}^{\mu}+A_{n}^{\mu}\Delta\tau$,
one finds

\begin{equation}
u_{n+1}^{\mu}u_{n+1,\mu}=1+\mathcal{O}(\Delta\tau^{2}),
\end{equation}
so that exact normalization is enforced by the renormalization step
above.

\paragraph{Strict causality and absence of runaways}

The update $A_{n}^{\mu}$ depends only on present and past information
($A_{\nu,n}^{L}$ and $A_{\nu,n-1}^{L}$). If $F^{\mu\nu}=0$, then
$A_{n}^{L}=0$ and $A_{n}^{\mu}=0$, so that $u_{n+1}^{\mu}=u_{n}^{\mu}$:
the particle moves inertially, with no self-excited runaway motion. 

\subsection{Energy--momentum balance}

Over one tick, the change in four-momentum $p^{\mu}=mu^{\mu}$ is

\begin{equation}
\Delta p_{n}^{\mu}=mA_{n}^{\mu}\Delta\tau.
\end{equation}
Substituting Eq. (\ref{eq:TTcovariant}), one obtains the identity

\[
\Delta p_{n}^{\mu}=qF^{\mu\nu}(x_{n})u_{\nu,n}\Delta\tau-R_{n}^{\mu}+\Delta S_{n}^{\mu}+\mathcal{O}(\Delta\tau^{2}),
\]
where

\begin{multline*}
R_{n}^{\mu}=-\,m\tau_{0}\,(A_{n}^{L}\!\cdot A_{n}^{L})\,u_{n}^{\mu}\,\Delta\tau,\\
\Delta S_{n}^{\mu}\equiv S_{n}^{\mu}-S_{n-1}^{\mu},\quad S_{n}^{\mu}=m\tau_{0}A_{L,n}^{\mu}.
\end{multline*}
Using

\[
u_{n}=u_{n-1}+A_{n-1}\,\Delta\tau+\mathcal{O}(\Delta\tau^{2})
\]
and $A_{n-1}\simeq A_{n-1}^{L}$ to leading order, one finds

\[
u_{n}\!\cdot\!A_{n-1}^{L}=A_{n-1}^{L}\!\cdot\!A_{n-1}^{L}\,\Delta\tau+\mathcal{O}(\Delta\tau^{2}),
\]
which yields the term $R_{n}^{\mu}$ shown below. Here $R_{n}^{\mu}$
represents the radiated four-momentum, with the minus sign ensuring
positive radiated energy since $A^{L}\!\cdot A^{L}<0$ for spacelike
four-acceleration, while $\Delta S_{n}^{\mu}$ is a Schott-like contribution.
Summing over ticks causes the Schott terms to telescope, leaving only
boundary contributions. Thus energy--momentum conservation is explicit,
with no hidden reservoir \citep{Sokolov2009,Sokolov2009_strongfield}. 

\subsection{Continuum limit }

Assuming smooth fields and taking $\Delta\tau\to0$, the finite difference
reduces to a derivative,

\begin{equation}
\frac{A_{\nu,n}^{L}-A_{\nu,n-1}^{L}}{\Delta\tau}\;\to\;\dot{A}_{\nu}^{L}.
\end{equation}
The TT update becomes
\begin{equation}
A^{\mu}=A_{L}^{\mu}+\tau_{0}\,\dot{A}_{L}^{\mu}+\mathcal{O}(\tau_{0}^{2}),
\end{equation}
which is precisely the Landau--Lifshitz four-acceleration. Thus,
in the continuum limit, TT dynamics reduces smoothly to LL, while
at finite $\Delta\tau$ it provides a causal, stable, and energy-consistent
update law.

\subsection{Connections to Other Discrete-Time Approaches\label{subsec:connections}}

The TT framework is conceptually related to earlier attempts to introduce
discreteness into the equations of motion, most notably the \emph{chronon model}
originally proposed by Caldirola \citep{Caldirola1956}. Both frameworks
share the idea that the motion of a charged particle proceeds through
finite proper-time intervals, yet they differ fundamentally in what
is discretized. The chronon theory acts on the kinematics---a finite-difference
equation for the four-velocity---and in the continuum limit reproduces
the higher-derivative ALD structure. TT dynamics, by contrast, discretizes
the \emph{field-driven acceleration}, ensuring orthogonality, causality,
and exact energy--momentum conservation at each step, and reduces
smoothly to the LL equation. 

This distinction illustrates a broader principle: discretization at
the level of the \emph{dynamical response}, rather than at the level
of the trajectory itself, can regularize self-interaction while preserving
physical consistency \citep{Spohn2004,Medina2006,Bauer2013}. 

From this perspective, TT dynamics may be viewed as a modern, energy-consistent
realization of ideas that motivated early discrete-time theories \citep{Caldirola1956,Krivitskii1991},
now formulated in a covariant and causal operator language. 

More generally, causal finite-step dynamics of this kind resonate
with discrete approaches to spacetime and field theory \citep{Bombelli1987,Oriti20092009},
where causality and conservation laws must remain compatible with
a fundamentally discrete evolution.

\section{Discussion and Implications\label{sec:discussion}}

The TT formulation provides a causal, deterministic, and energy-consistent
alternative to both the ALD and LL equations. Its main advantages
can be summarized as follows:

\paragraph{Resolution of ALD pathologies.---}

By construction, TT dynamics avoids both runaways and pre-acceleration.
The update law (\ref{eq:TTupdate}) depends only on present and past
ticks, ensuring \emph{strict causality}. In the absence of external
fields ($F^{\mu\nu}=0$), the update reduces to uniform motion, so
that no self-excited acceleration can occur. This directly resolves
the longstanding instabilities of the ALD equation \citep{Rohrlich2007,Spohn2004}.

\paragraph{Clarification of Landau--Lifshitz foundations.---}

The LL equation has been remarkably successful in practice, but it
is derived by a reduction-of-order approximation whose physical mechanism
is obscured by the Schott energy reservoir. TT dynamics, in contrast,
provides a discrete stepwise mechanism in which energy accounting
is explicit. Equation (\ref{eq:TTenergy}) shows that the external
work divides cleanly into Larmor radiation loss plus a Schott-like
boundary term. The latter telescopes across ticks, eliminating the
need for a hidden energy reservoir. In this way, TT dynamics places
the LL equation on a transparent and physically grounded foundation.

\paragraph{Natural high-frequency cutoff.---}

In frequency space, the TT formulation multiplies the radiation--reaction
contribution by

\[
e^{-i\omega\Delta t/2}\,\operatorname{sinc}\!\left(\frac{\omega\Delta t}{2}\right),
\]
ensuring agreement with the LL limit in the smooth-field regime $\omega\Delta t\ll1$,
while automatically damping high-frequency components near $|\omega\Delta t|\sim\pi$.
This finite-tick factor introduces both a causal half-tick retardation
and a smooth attenuation, which together suppress unphysical high-frequency
growth without altering the correct low-frequency behavior. This property
is absent in the LL equation and is of practical importance in numerical
simulations of radiation reaction in laser--plasma interactions \citep{DiPiazza2012,Gonoskov2022,Kirk2009,Liseykina2016}.
It also provides a natural regulator for discontinuous or rapidly
varying fields, achieved without any ad~hoc assumptions.

\paragraph{Connection to characteristic timescales.---}

The finite tick size $\Delta t$ is not fixed \emph{a priori}. Identifying
it with the radiation--reaction timescale $\tau_{0}$ {[}Eq. (\ref{eq:tau0}){]}
yields a natural causal cutoff at frequencies $\omega\sim\pi/\tau_{0}$.
For the electron, this corresponds to photon energies of order hundreds
of MeV, which coincides with the onset of quantum electrodynamical
(QED) corrections \citep{Higuchi2002,Higuchi2006,Dunne2009,DiPiazza2012}.
This suggests that TT dynamics may capture the classical limit of
radiation reaction up to the very boundary where quantum effects become
essential. 

\paragraph{Implications for quantum extensions.---}

Because TT dynamics provides a stable and causal baseline, it offers
a natural platform for quantum generalizations. Quantum corrections
such as radiation quenching \citep{Harvey2017}, stochastic fluctuations
\citep{Torgrimsson2020}, and plasma-field modifications \citep{Ilderton2016,Niel2018,Gonoskov2022}
can be interpreted relative to a consistent classical foundation.
In this sense, TT dynamics not only clarifies the classical problem
but also bridges toward a deeper understanding of radiation reaction
in strong-field QED. 

\paragraph{Broader perspective.---}

From a broader viewpoint, TT dynamics achieves what has long been
sought: a simple update law that eliminates pathologies, ensures exact
tick-wise energy--momentum conservation, and reduces to LL in the
continuum limit. It thus provides both conceptual clarity and practical
stability, and may find applications in diverse contexts ranging from
plasma physics to high-intensity laser experiments, and potentially
to effective models of electron dynamics at the boundary between classical
and quantum regimes.

\section{Outlook and Future Work\label{sec:Outlook}}

The TT framework suggests several promising directions for further
exploration, both conceptually and practically.

\paragraph{Fundamental interpretation of $\Delta t$.---}

A central open question is the physical meaning of the tick size $\Delta t$.
While identifying $\Delta t$ with the radiation--reaction timescale
$\tau_{0}$ yields a natural cutoff consistent with the onset of QED
effects, it remains to be understood whether this identification is
unique or emergent from a deeper principle. Future work may investigate
whether $\Delta t$ arises naturally in effective field theory or
in semiclassical limits of QED.

\paragraph{Numerical implementations.---}

Because TT dynamics is explicit, causal, and free of pathologies,
it is well suited for numerical simulations of charged-particle dynamics
in strong fields. The finite-difference formulation automatically
regularizes high-frequency modes, which may improve the stability
and accuracy of particle-in-cell (PIC) algorithms used in plasma and
laser--matter interaction studies \citep{Kirk2009,Liseykina2016}.
Benchmarking TT-based simulations against LL-based methods and experimental
results will be a valuable next step \citep{Gonoskov2015,Vranic2016,Sokolov2009,Naumova2009,Capdessus2015}.

\paragraph{Extensions to quantum regimes.---}

In ultra-intense laser facilities, radiation reaction enters a regime
where quantum effects such as stochastic photon emission, radiation
quenching, and pair production become important \citep{DiPiazza2012,Harvey2017}.
Extending TT dynamics to include quantum corrections is a natural
avenue of research. Because TT provides a stable and causal baseline,
one can systematically add stochastic or quantum corrections as modifications
to the tick update rule, potentially yielding a hybrid classical--quantum
algorithm \citep{DiPiazza2008,Ilderton2019}.

\paragraph{Connections to other discrete-time approaches.---}

TT dynamics is related in spirit to earlier discrete-time proposals,
such as Caldirola's chronon theory \citep{Caldirola1956}, but with
transparent energy--momentum bookkeeping. Exploring connections to
modern discrete approaches in quantum foundations, causal set theory,
and lattice gauge theory may reveal deeper structural links. In particular,
the telescoping Schott term and causal cutoff may have analogues in
other discrete spacetime frameworks.

\paragraph{Experimental prospects.---}

Although direct tests of radiation reaction remain challenging, upcoming
ultra-intense laser experiments are expected to probe regimes where
classical and quantum recoil compete \citep{Cole2018,Poder2018}.
The TT framework suggests concrete predictions, such as the suppression
of high-frequency spectral components relative to LL theory, which
could be experimentally accessible. Identifying measurable observables
that distinguish TT dynamics from standard LL predictions will be
an important task.

\paragraph{Broader implications.---}

Finally, TT dynamics illustrates how finite-step formulations can
resolve pathologies of higher-derivative theories while maintaining
consistency with known physics in the continuum limit. This perspective
may prove useful beyond electrodynamics, for example in radiation
reaction of gravitational systems, effective theories of dissipative
forces, or other contexts where self-interaction leads to divergences.
\medskip In summary, TT dynamics provides a stable and causal resolution
of the classical radiation--reaction problem while opening new avenues
for theoretical and experimental investigation. Its unification of
energy transparency, causal stability, and a natural cutoff suggests
that tick-based formulations may offer fresh insight into the interface
between classical and quantum dynamics.

\bibliographystyle{elsarticle-num}
\bibliography{export}

\begin{thebibliography}{10}
\expandafter\ifx\csname url\endcsname\relax
  \def\url#1{\texttt{#1}}\fi
\expandafter\ifx\csname urlprefix\endcsname\relax\def\urlprefix{URL }\fi
\expandafter\ifx\csname href\endcsname\relax
  \def\href#1#2{#2} \def\path#1{#1}\fi

\bibitem{PAMDirac1938}
P.~A.~M. Dirac, Classical theory of radiating electrons, Proceedings of the Royal Society of London. Series A. Mathematical and Physical Sciences 167 (1938) 148--169.
\newblock \href {https://doi.org/10.1098/rspa.1938.0124} {\path{doi:10.1098/rspa.1938.0124}}.

\bibitem{Rohrlich2007}
F.~Rohrlich, Classical Charged Particles, WORLD SCIENTIFIC, 2007.
\newblock \href {https://doi.org/10.1142/6220} {\path{doi:10.1142/6220}}.

\bibitem{Griffiths2012}
D.~J. Griffiths, Resource letter em-1: Electromagnetic momentum, American Journal of Physics 80 (2012) 7--18.
\newblock \href {https://doi.org/10.1119/1.3641979} {\path{doi:10.1119/1.3641979}}.

\bibitem{Klepikov1985}
N.~P. Klepikov, Radiation damping forces and radiation from charged particles, Soviet Physics Uspekhi 28 (1985) 506--520.
\newblock \href {https://doi.org/10.1070/PU1985v028n06ABEH005205} {\path{doi:10.1070/PU1985v028n06ABEH005205}}.

\bibitem{Spohn2004}
H.~Spohn, Dynamics of Charged Particles and their Radiation Field, Cambridge University Press, 2004.
\newblock \href {https://doi.org/10.1017/CBO9780511535178} {\path{doi:10.1017/CBO9780511535178}}.

\bibitem{Landau1989}
L.~D. Landau, E.~M. Lifshitz, The classical theory of fields, Pergamon Press, 1989.

\bibitem{DiPiazza2012}
A.~D. Piazza, C.~M\"uller, K.~Z. Hatsagortsyan, C.~H. Keitel, Extremely high-intensity laser interactions with fundamental quantum systems, Reviews of Modern Physics 84 (2012) 1177--1228.
\newblock \href {https://doi.org/10.1103/RevModPhys.84.1177} {\path{doi:10.1103/RevModPhys.84.1177}}.

\bibitem{Gonoskov2022}
A.~Gonoskov, T.~Blackburn, M.~Marklund, S.~Bulanov, Charged particle motion and radiation in strong electromagnetic fields, Reviews of Modern Physics 94 (2022) 045001.
\newblock \href {https://doi.org/10.1103/RevModPhys.94.045001} {\path{doi:10.1103/RevModPhys.94.045001}}.

\bibitem{Kravets2013}
Y.~Kravets, A.~Noble, D.~Jaroszynski, Radiation reaction effects on the interaction of an electron with an intense laser pulse, Physical Review E 88 (2013) 011201.
\newblock \href {https://doi.org/10.1103/PhysRevE.88.011201} {\path{doi:10.1103/PhysRevE.88.011201}}.

\bibitem{Spohn2000}
H.~Spohn, The critical manifold of the lorentz-dirac equation, Europhysics Letters (EPL) 50 (2000) 287--292.
\newblock \href {https://doi.org/10.1209/epl/i2000-00268-x} {\path{doi:10.1209/epl/i2000-00268-x}}.

\bibitem{Ford1991}
G.~Ford, R.~O'Connell, Radiation reaction in electrodynamics and the elimination of runaway solutions, Physics Letters A 157 (1991) 217--220.
\newblock \href {https://doi.org/10.1016/0375-9601(91)90054-C} {\path{doi:10.1016/0375-9601(91)90054-C}}.

\bibitem{Ford1993}
G.~Ford, R.~O'Connell, Relativistic form of radiation reaction, Physics Letters A 174 (1993) 182--184.
\newblock \href {https://doi.org/10.1016/0375-9601(93)90755-O} {\path{doi:10.1016/0375-9601(93)90755-O}}.

\bibitem{Medina2006}
R.~Medina, Radiation reaction of a classical quasi-rigid extended particle, Journal of Physics A: Mathematical and General 39 (2006) 3801--3816.
\newblock \href {https://doi.org/10.1088/0305-4470/39/14/021} {\path{doi:10.1088/0305-4470/39/14/021}}.

\bibitem{Bauer2013}
G.~Bauer, D.-A. Deckert, D.~D\"urr, Maxwell-lorentz dynamics of rigid charges, Communications in Partial Differential Equations 38 (2013) 1519--1538.
\newblock \href {https://doi.org/10.1080/03605302.2013.814142} {\path{doi:10.1080/03605302.2013.814142}}.

\bibitem{Caldirola1956}
P.~Caldirola, A new model of classical electron, Il Nuovo Cimento 3 (1956) 297--343.
\newblock \href {https://doi.org/10.1007/BF02743686} {\path{doi:10.1007/BF02743686}}.

\bibitem{Krivitskii1991}
V.~S. Krivitskii, V.~N. Tsytovich, Average radiation-reaction force in quantum electrodynamics, Soviet Physics Uspekhi 34 (1991) 250--258.
\newblock \href {https://doi.org/10.1070/PU1991v034n03ABEH002352} {\path{doi:10.1070/PU1991v034n03ABEH002352}}.

\bibitem{Sokolov2009_strongfield}
I.~V. Sokolov, N.~M. Naumova, J.~A. Nees, G.~A. Mourou, V.~P. Yanovsky, Dynamics of emitting electrons in strong laser fields, Physics of Plasmas 16 (9 2009).
\newblock \href {https://doi.org/10.1063/1.3236748} {\path{doi:10.1063/1.3236748}}.

\bibitem{DiPiazza2008}
A.~D. Piazza, Exact solution of the landau-lifshitz equation in a plane wave, Letters in Mathematical Physics 83 (2008) 305--313.
\newblock \href {https://doi.org/10.1007/s11005-008-0228-9} {\path{doi:10.1007/s11005-008-0228-9}}.

\bibitem{Ilderton2013}
A.~Ilderton, G.~Torgrimsson, Radiation reaction in strong field qed, Physics Letters B 725 (2013) 481--486.
\newblock \href {https://doi.org/10.1016/j.physletb.2013.07.045} {\path{doi:10.1016/j.physletb.2013.07.045}}.

\bibitem{Torgrimsson2020}
G.~Torgrimsson, Nonlinear photon trident versus double compton scattering and resummation of one-step terms, Physical Review D 102 (2020) 116008.
\newblock \href {https://doi.org/10.1103/PhysRevD.102.116008} {\path{doi:10.1103/PhysRevD.102.116008}}.

\bibitem{Higuchi2002}
A.~Higuchi, Radiation reaction in quantum field theory, Physical Review D 66 (2002) 105004.
\newblock \href {https://doi.org/10.1103/PhysRevD.66.105004} {\path{doi:10.1103/PhysRevD.66.105004}}.

\bibitem{Niel2018}
F.~Niel, C.~Riconda, F.~Amiranoff, R.~Duclous, M.~Grech, From quantum to classical modeling of radiation reaction: A focus on stochasticity effects, Physical Review E 97 (2018) 043209.
\newblock \href {https://doi.org/10.1103/PhysRevE.97.043209} {\path{doi:10.1103/PhysRevE.97.043209}}.

\bibitem{Dunne2009}
G.~V. Dunne, New strong-field qed effects at extreme light infrastructure, The European Physical Journal D 55 (2009) 327--340.
\newblock \href {https://doi.org/10.1140/epjd/e2009-00022-0} {\path{doi:10.1140/epjd/e2009-00022-0}}.

\bibitem{Jackson2021}
J.~D. Jackson, Classical Electrodynamics, International Adaptation, John Wiley \& Sons, Incorporated, 2021.

\bibitem{Gralla2009}
S.~E. Gralla, A.~I. Harte, R.~M. Wald, Rigorous derivation of electromagnetic self-force, Physical Review D 80 (2009) 024031.
\newblock \href {https://doi.org/10.1103/PhysRevD.80.024031} {\path{doi:10.1103/PhysRevD.80.024031}}.

\bibitem{Bild2019}
C.~Bild, D.-A. Deckert, H.~Ruhl, Radiation reaction in classical electrodynamics, Physical Review D 99 (2019) 096001.
\newblock \href {https://doi.org/10.1103/PhysRevD.99.096001} {\path{doi:10.1103/PhysRevD.99.096001}}.

\bibitem{Kirk2009}
J.~G. Kirk, A.~R. Bell, I.~Arka, Pair production in counter-propagating laser beams, Plasma Physics and Controlled Fusion 51 (2009) 085008.
\newblock \href {https://doi.org/10.1088/0741-3335/51/8/085008} {\path{doi:10.1088/0741-3335/51/8/085008}}.

\bibitem{Yaghjian1992}
A.~D. Yaghjian, Relativistic Dynamics of a Charged Sphere, Vol.~11, Springer New York, 1992.
\newblock \href {https://doi.org/10.1007/978-0-387-73967-0} {\path{doi:10.1007/978-0-387-73967-0}}.

\bibitem{Liseykina2016}
T.~V. Liseykina, S.~V. Popruzhenko, A.~Macchi, Inverse faraday effect driven by radiation friction, New Journal of Physics 18 (2016) 072001.
\newblock \href {https://doi.org/10.1088/1367-2630/18/7/072001} {\path{doi:10.1088/1367-2630/18/7/072001}}.

\bibitem{Burton2014}
D.~A. Burton, A.~Noble, Aspects of electromagnetic radiation reaction in strong fields, Contemporary Physics 55 (2014) 110--121.
\newblock \href {https://doi.org/10.1080/00107514.2014.886840} {\path{doi:10.1080/00107514.2014.886840}}.

\bibitem{Baylis2002}
W.~Baylis, J.~Huschilt, Energy balance with the landau--lifshitz equation, Physics Letters A 301 (2002) 7--12.
\newblock \href {https://doi.org/10.1016/S0375-9601(02)00963-5} {\path{doi:10.1016/S0375-9601(02)00963-5}}.

\bibitem{Sokolov2009}
I.~V. Sokolov, Renormalization of the lorentz-abraham-dirac equation for radiation reaction force in classical electrodynamics, Journal of Experimental and Theoretical Physics 109 (2009) 207--212.
\newblock \href {https://doi.org/10.1134/S1063776109080044} {\path{doi:10.1134/S1063776109080044}}.

\bibitem{Bombelli1987}
L.~Bombelli, J.~Lee, D.~Meyer, R.~D. Sorkin, Space-time as a causal set, Physical Review Letters 59 (1987) 521--524.
\newblock \href {https://doi.org/10.1103/PhysRevLett.59.521} {\path{doi:10.1103/PhysRevLett.59.521}}.

\bibitem{Oriti20092009}
D.~Oriti, Approaches to Quantum Gravity, Cambridge University Press, 2009.
\newblock \href {https://doi.org/10.1017/CBO9780511575549} {\path{doi:10.1017/CBO9780511575549}}.

\bibitem{Higuchi2006}
A.~Higuchi, G.~D.~R. Martin, Radiation reaction on charged particles in three-dimensional motion in classical and quantum electrodynamics, Physical Review D 73 (2006) 025019.
\newblock \href {https://doi.org/10.1103/PhysRevD.73.025019} {\path{doi:10.1103/PhysRevD.73.025019}}.

\bibitem{Harvey2017}
C.~Harvey, A.~Gonoskov, A.~Ilderton, M.~Marklund, Quantum quenching of radiation losses in short laser pulses, Physical Review Letters 118 (2017) 105004.
\newblock \href {https://doi.org/10.1103/PhysRevLett.118.105004} {\path{doi:10.1103/PhysRevLett.118.105004}}.

\bibitem{Ilderton2016}
A.~Ilderton, M.~Marklund, Prospects for studying vacuum polarisation using dipole and synchrotron radiation, Journal of Plasma Physics 82 (2016) 655820201.
\newblock \href {https://doi.org/10.1017/S0022377816000192} {\path{doi:10.1017/S0022377816000192}}.

\bibitem{Gonoskov2015}
A.~Gonoskov, S.~Bastrakov, E.~Efimenko, A.~Ilderton, M.~Marklund, I.~Meyerov, A.~Muraviev, A.~Sergeev, I.~Surmin, E.~Wallin, Extended particle-in-cell schemes for physics in ultrastrong laser fields: Review and developments, Physical Review E 92 (2015) 023305.
\newblock \href {https://doi.org/10.1103/PhysRevE.92.023305} {\path{doi:10.1103/PhysRevE.92.023305}}.

\bibitem{Vranic2016}
M.~Vranic, T.~Grismayer, R.~A. Fonseca, L.~O. Silva, Quantum radiation reaction in head-on laser-electron beam interaction, New Journal of Physics 18 (2016) 073035.
\newblock \href {https://doi.org/10.1088/1367-2630/18/7/073035} {\path{doi:10.1088/1367-2630/18/7/073035}}.

\bibitem{Naumova2009}
N.~Naumova, T.~Schlegel, V.~T. Tikhonchuk, C.~Labaune, I.~V. Sokolov, G.~Mourou, Hole boring in a dt pellet and fast-ion ignition with ultraintense laser pulses, Physical Review Letters 102 (2009) 025002.
\newblock \href {https://doi.org/10.1103/PhysRevLett.102.025002} {\path{doi:10.1103/PhysRevLett.102.025002}}.

\bibitem{Capdessus2015}
R.~Capdessus, P.~McKenna, Influence of radiation reaction force on ultraintense laser-driven ion acceleration, Physical Review E 91 (2015) 053105.
\newblock \href {https://doi.org/10.1103/PhysRevE.91.053105} {\path{doi:10.1103/PhysRevE.91.053105}}.

\bibitem{Ilderton2019}
A.~Ilderton, B.~King, D.~Seipt, Extended locally constant field approximation for nonlinear compton scattering, Physical Review A 99 (2019) 042121.
\newblock \href {https://doi.org/10.1103/PhysRevA.99.042121} {\path{doi:10.1103/PhysRevA.99.042121}}.

\bibitem{Cole2018}
J.~Cole, K.~Behm, E.~Gerstmayr, T.~Blackburn, J.~Wood, C.~Baird, M.~Duff, C.~Harvey, A.~Ilderton, A.~Joglekar, K.~Krushelnick, S.~Kuschel, M.~Marklund, P.~McKenna, C.~Murphy, K.~Poder, C.~Ridgers, G.~Samarin, G.~Sarri, D.~Symes, A.~Thomas, J.~Warwick, M.~Zepf, Z.~Najmudin, S.~Mangles, Experimental evidence of radiation reaction in the collision of a high-intensity laser pulse with a laser-wakefield accelerated electron beam, Physical Review X 8 (2018) 011020.
\newblock \href {https://doi.org/10.1103/PhysRevX.8.011020} {\path{doi:10.1103/PhysRevX.8.011020}}.

\bibitem{Poder2018}
K.~Poder, M.~Tamburini, G.~Sarri, A.~D. Piazza, S.~Kuschel, C.~Baird, K.~Behm, S.~Bohlen, J.~Cole, D.~Corvan, M.~Duff, E.~Gerstmayr, C.~Keitel, K.~Krushelnick, S.~Mangles, P.~McKenna, C.~Murphy, Z.~Najmudin, C.~Ridgers, G.~Samarin, D.~Symes, A.~Thomas, J.~Warwick, M.~Zepf, Experimental signatures of the quantum nature of radiation reaction in the field of an ultraintense laser, Physical Review X 8 (2018) 031004.
\newblock \href {https://doi.org/10.1103/PhysRevX.8.031004} {\path{doi:10.1103/PhysRevX.8.031004}}.

\end{thebibliography}

\end{document}